# The local environmental dependence of galaxy properties in the volume-limited sample of Main Galaxies from the SDSS Data Release 5

Xin-Fa Deng    Ji-Zhou He    Qun Zhang    Cong-Gen He    Peng Jiang    Yong Xin

School of Science, Nanchang University, Jiangxi, China, 330047

**Abstract**   Using a volume-limited sample of Main Galaxies from the SDSS Data Release 5, we investigate the dependence of galaxy properties on local environment. For each galaxy, the local three-dimensional density is calculated. We find that galaxy morphologies strongly depend on local environment: galaxies in dense environments have predominantly early type morphologies, but other galaxy properties do not present significant dependence on local environment. Clearly, this puts a important constraint on proposed physical mechanisms.

## 1. Introduction

The numerous studies showed that galaxy properties seem to correlate (significantly) with environment, for example, galaxies in dense environments (i.e., clusters or groups) have high proportion of early type morphologies (e.g., Oemler 1974; Dressler 1980; Whitmore, Gilmore & Jones 1993, Deng et al. 2006a) and low SFRs (e.g., Balogh et al. 1997, 1999; Poggianti et al. 1999). Many authors have investigated correlations between environment and galaxy properties, such as ones between environment and morphology (e.g., Postman & Geller 1984; Dressler et al. 1997; Hashimoto & Oemler 1999; Fasano et al. 2000; Tran et al. 2001; Goto et al. 2003; Helsdon & Ponman 2003; Treu et al. 2003), ones between environment and star formation rate (e.g., Hashimoto et al. 1998; Lewis et al. 2002; G´omez et al. 2003; Balogh et al. 2004a; Tanaka et al. 2004; Kelm, Focardi & Sorrentino 2005), and ones between environment and colour (e.g.,Tanaka et al. 2004; Balogh et al. 2004b; Hogg et al. 2004).   In order to explain these correlations, various physical mechanisms have been proposed, including rampressure stripping (Gunn & Gott 1972; Kent 1981; Fujita & Nagashima 1999; Quilis, Moore & Bower 2000); galaxy harassment (Moore et al. 1996, 1999); cluster tidal forces (Byrd &Valtonen 1990; Valluri 1993; Gnedin 2003); and interaction/merging of galaxies (Icke 1985; Lavery & Henry 1988; Mamon 1992; Bekki 1998).

In order to study the morphology-density relation and morphology–cluster-centric-radius relation, Goto et al.(2003) measured local galaxy density in the following way. For each galaxy, they calculated the projected distance to the 5th nearest galaxy within ±1000 km s$^{-1}$ in redshift space in a volume limited sample (0.05< z <0.1, $M_r$ < −20.5) of the Sloan Digital Sky Survey (SDSS) data. The local galaxy density was defined as the number of galaxies (N = 5) within the distance to the circular surface area with the radius of the distance to the 5th nearest galaxy. They found that the morphology-density and morphology–cluster-centric-radius relation as fractions of early-type galaxies increase and as those of late-type galaxies decrease toward increasing local galaxy density. In addition, they also found there are two characteristic changes in both the morphology-density and the morphology-radius relations, suggesting two different mechanisms are responsible for the relations.

If redshift is taken as a pure distance measure, above local galaxy density is actually thequasi-three-dimensional density. In this paper, we modify Goto et al.(2003)'s method and measure local three-dimensional galaxy density in the following way. For each galaxy, we calculate the three-dimensional distance to the 5th nearest galaxy. The local three-dimensional galaxy density is defined as the number of galaxies (N=5) within this distance to the the volume of the sphere with the radius of the distance to the 5th nearest galaxy. We will all-roundly investigate correlations between environment and galaxy properties. Our paper is organized as follows. In section 2, we describe the data to be used. The correlations between galaxy properties and local galaxy density are discussed in section 3. Our main results and conclusions are summarized in section 4.

**2. Data**

The Sloan Digital Sky Survey (SDSS) is one of the largest astronomical surveys to date. The completed survey will cover approximately 10000 square degrees. York et al. (2000) provided the technical summary of the SDSS. The SDSS observes galaxies in five photometric bands (u, g, r, i, z) centered at (3540, 4770, 6230, 7630, 9130 Å). The imaging camera was described by Gunn et al. (1998), while the photometric system and the photometric calibration of the SDSS imaging data were roughly described by Fukugita et al. (1996), Hogg et al. (2001) and Smith et al. (2002) respectively. Pier et al. (2003) described the methods and algorithms involved in the astrometric calibration of the survey, and present a detailed analysis of the accuracy achieved. Many of the survey properties were discussed in detail in the Early Data Release paper (Stoughton et al. 2002). Galaxy spectroscopic target selection can be implemented by two algorithms. The MAIN Galaxy sample (Strauss et al. 2002) targets galaxies brighter than $r_p < 17.77$(r-band apparent Petrosian magnitude). Most galaxies of this sample are within redshift region $0.02 \leq z \leq 0.2$. The Luminous Red Galaxy (LRG) algorithm (Eisenstein et al. 2001) selects galaxies to $r_p < 19.5$ that are likely to be luminous early-types, using color-magnitude cuts in g, r, and i. Because most LRGs are within redshift region $0.2 \leq z \leq 0.4$, two samples mentioned above actually represent the distribution of galaxies located at different depth.

The SDSS has adopted a modified form of the Petrosian (1976) system for galaxy photometry. The Petrosian radius $r_p$ is defined to be the radius at which the local surface-brightness averaged in an annulus $0.8r_p$ -$1.25r_p$ equals 20% of the mean surface-brightness within radius $r_p$:

$$\frac{\int_{0.8r_p}^{1.25r_p} dr 2\pi \, rI(r)/[\pi(1.25^2 - 0.8^2)r^2]}{\int_0^{r_p} dr 2\pi \, rI(r)/[\pi r^2]} = 0.2$$

where I(r) is the azimuthally averaged surface-brightness profile. The Petrosian flux $F_p$ in any band is then defined as the total flux within a radius of $2r_p$: $F_p = \int_0^{2r_p} 2\pi\, rdr I(r)$. The aperture $2r_p$ is large enough to contain nearly all of the flux for typical galaxy profiles, but small enough that the sky noise in $F_p$ is small. Theoretically, the Petrosian flux (magnitude) defined here should recover about 98 % of the total flux for an exponential profile and about 80% for a de Vaucouleurs profile. The other two Petrosian radii listed in the Photo output, $R_{50}$ and $R_{90}$, are the radii enclosing 50% and 90% of the Petrosian flux, respectively.

In our work, we consider the Main galaxy sample. The data is download from the Catalog Archive Server of SDSS Data Release 5 by the SDSS SQL Search (with SDSS flag: bestPrimtarget=64) with high-confidence redshifts ($Zwarning \neq 16$ and $Zstatus \neq 0, 1$ and redshift confidence level: zconf>0.95) (http://www.sdss.org/dr5/). From this sample, we select 332412 Main galaxies in redshift region: $0.02 \leq z \leq 0.2$.

In calculating the distance we use a cosmological model with a matter density $\Omega_0 = 0.3$, cosmological constant $\Omega_A = 0.7$, Hubble's constant $H_0 = 100 h\,\text{km}\cdot\text{s}^{-1}\cdot\text{Mpc}^{-1}$ with h=0.7.

We intend to construct a volume-limited sample from the Main galaxy sample of SDSS5. This choice reduces the number of galaxies available, but it has several important advantages: the radial selection function is approximately uniform, thus the only variation in the space density of galaxies with radial distance is due to clustering. Thus, the analysis is more straight forward. The volume-limited sample is defined by choosing minimum and maximum absolute magnitude limits. According to the distribution of absolute magnitude for the Main galaxy sample, we choose the maximum absolute magnitude limit $M_{max} = -22.40$. The absolute magnitude M is the r-band absolute magnitude. In our work, we ignore the K-correction (Blanton et al. 2003a). The minimum absolute magnitude limit is defined as the absolute magnitude of a galaxy having r-band apparent Petrosian magnitude r=17.77 ( the r-band apparent Petrosian magnitude limit of the Main galaxy sample) at a redshift limit $Z_{max}$. Figure1 shows the number of galaxies in the volume-limited sample as a function of redshift limit $Z_{max}$. As seen from this figure, there is the highest platform in the redshift region $0.09 \leq z \leq 0.124$. When constructing a volume-limited sample, we mainly consider two factors: the luminosity region of the volume-limited sample is as large as possible, and the galaxy number of the sample as many as possible. Thus, we construct a volume-limited sample that extends to $Z_{max} = 0.09$, and limits the absolute magnitude region: $-22.40 \leq M_r \leq -20.30$, which contains 69381 galaxies, and in which the mean galaxy density is about $2.2386 \times 10^{-3}\, Mpc^{-3}$. Figure 2 shows the proportion of early-type galaxies in different luminosity bins (bin $\Delta M_r = 0.4$) for the whole Main galaxy sample. We note that from about $M_r = -20.20$ the proportion of early-type galaxies rapidly increase with increasing luminosity. So, this volume-limited sample is also a good sample for investigating the correlation between galaxy morphological types and luminosities.

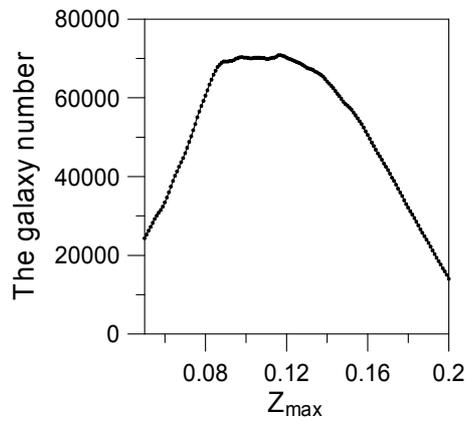

Fig.1    The number of galaxies in a volume-limited sample as a function of redshift limit $z_{max}$.

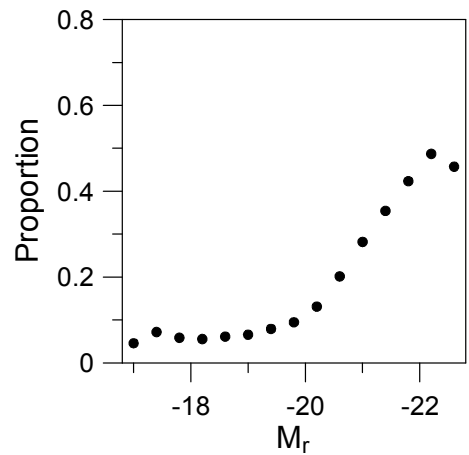

Fig.2    The proportion of early-type galaxies in different luminosity bins for the whole Main galaxy sample.

## 3. Correlations between galaxy properties and local galaxy density

For each galaxy, we calculate the three-dimensional distance to the 5th nearest galaxy. The local three-dimensional galaxy density is defined as the number of galaxies (N=5) within this distance to the the volume of the sphere with the radius of the distance to the 5th nearest galaxy. In this paper, we express the local galaxy density in relative density (the ratio of the local three-dimensional density of galaxies to the mean galaxy density of the volume-limited sample). Figure 3 shows the distribution of the local relative density for all galaxies.

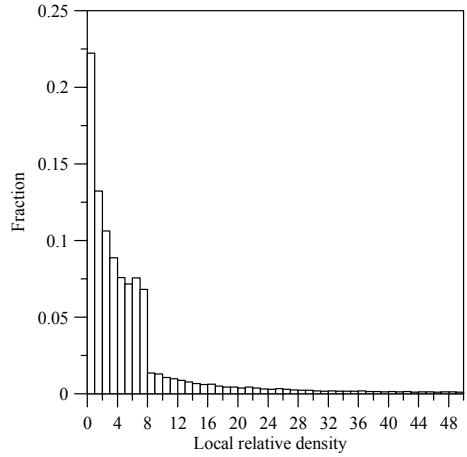

Fig.3 Distribution of the local relative density(within 5 th) for all galaxies.

In this paper, the concentration index $c_i = R_{90}/R_{50}$ is used to separate early-type (E/S0) galaxies from late-type (Sa/b/c, Irr) galaxies. (Shimasaku et al. 2001). Using about 1500 galaxies with eye-ball classification, Nakamura et al. (2003) confirmed that $c_i = 2.86$ separates galaxies at S0/a with a completeness of about 0.82 for both late and early types. We calculate the proportion of early-type ($c_i \geq 2.86$) galaxies in different density bins (bin=1). Figure 4 illustrates this proportion as a function of the local relative density of galaxies. The dashed line represents the proportion of early-type galaxies of the volume-limited sample. As seen from this figure, although there is larger scatter in high density region (it can be caused by fewer galaxies with high local density), the proportion of early type galaxies clearly increases with increasing density. Many previous works showed galaxies in dense environments (i.e., clusters or groups) have high proportion of early type morphologies (e.g., Oemler 1974; Dressler 1980; Whitmore, Gilmore & Jones 1993, Deng et al. 2006a), while galaxies in the lowest density regions (isolated galaxies) have lower proportion of early-type galaxies (e.g., Deng et al. 2006b). Our alyses further confirm this correlation between morphological types of galaxies and local galaxy density.

Applying the projected correlation functions $w_p(r_p)$, Zehavi et al.(2002) found that more luminous galaxies strongly cluster. Using photometry and spectroscopy of 144,609 galaxies from the Sloan Digital Sky Survey, Blanton et al. (2003b) investigated the dependence of local galaxy density (smoothed on 8 $h^{-1}$ Mpc scales) on seven galaxy properties: four optical colors, surface brightness, radial profile shape as measured by the Sérsic index, and absolute magnitude. They found that local density is a strong function of luminosity, and the most luminous galaxies exist preferentially in the densest regions of the universe. These results were similar to those found in a number of earlier studies (Davis et al. 1988; Hamilton 1988; Park et al. 1994; Loveday et al. 1995; Guzzo et al. 1997; Benoist et al. 1998; Norberg et al. 2001). In figure 5, we present the mean luminosity as a function of the local relative density of galaxies. Error bars are standard deviation in each density bin. As seen from this figure, we do not observe any correlation between galaxy luminosity and local density. This result agrees with that found by Deng et al. (2006a). By comparing statistical properties of galaxy luminosity in the compact galaxy group sample with those in random group sample, Deng et al. (2006a) found that the two samples have the same statistical properties of luminosity.

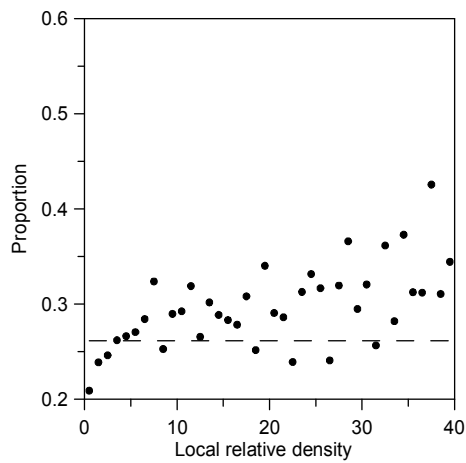

Fig.4 Proportion of early-type galaxies as a function of the local relative density(within 5th) of galaxies. The dashed line represents the proportion of early-type galaxies of the volume-limited sample.

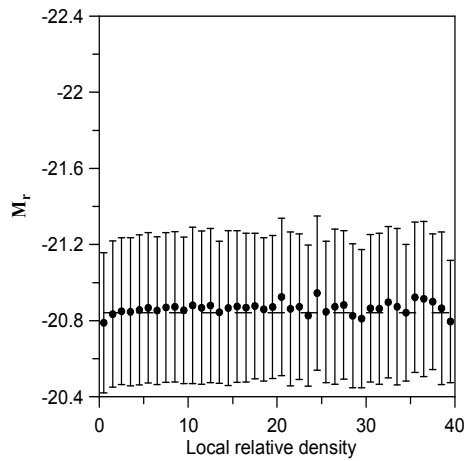

Fig.5 Mean luminosity as a function of the local relative density (within 5th) of galaxies. The dashed line represents the mean luminosity of the volume-limited sample. Error bars are standard deviation in each density bin.

Norberg et al. (2002) investigated the dependence of galaxy clustering on luminosity and spectral type using the 2dF Galaxy Redshift Survey (2dFGRS). The galaxy sample were divided into two broad spectral classes: galaxies with strong emission lines ('late-types'), and more quiescent galaxies ('early-types'). They calculated the projected correlation functions of both spectral types, and found that both early and late types have approximately the same dependence of clustering strength on luminosity.These results demonstrated that luminosity, and not type, is the dominant factor of galaxy clustering. According to above analyses, our results show that morphological types of galaxies are strongly correlated with local density, while there is no significant correlation between galaxy luminosity and local density. Apparently, these results do not agree with those found by Norberg et al. (2002).

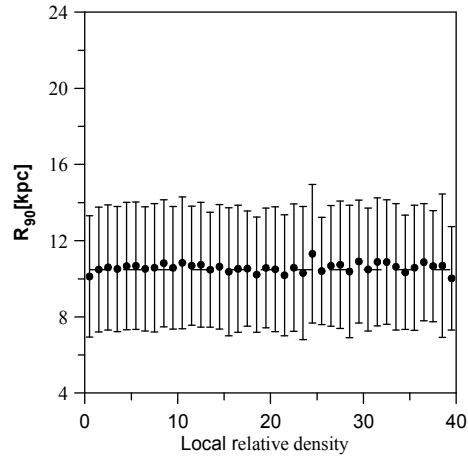

Fig.6 Mean size as a function of the local relative density(within 5th) of galaxies. The dashed line represents the mean size of the volume-limited sample. Error bars are standard deviation in each density bin.

We also present the mean size as a function of the local relative density of galaxies. The r-band $R_{90,r}$ is selected as the parameter of galaxy size. As seen from figure 6, there is no significant correlation between galaxy size and local density.

Galaxy colors are an important quantity that characterizes stellar contents of galaxies. Some studies showed that clustering of galaxies depends on color (Brown et al. 2000; Zehavi et al. 2002). Blanton et al. (2003b) also found that local density is a strong function of all colors. In this paper, we intend to investigate the correlation of color with local environment. Figure 7-10 present four optical mean colors as a function of the local relative density. As seen in these figures, except g-r color, we only observe a weak dependence of color on local density. This result is similar to that found by other authors (Bernardi et al. 2003; Balogh et al. 2004b; Hogg et al. 2004). Hogg et al. (2004)' study showed that although the most luminous galaxies reside preferentially in the highest density regions, red galaxy colors are independence on environment.  By analysing the $u - r$ color distribution of galaxies as a function of luminosity and environment, Balogh et al. (2004b) found that at fixed luminosity the mean color of blue galaxies or red galaxies is nearly independent of environment, in contrast, at fixed luminosity the fraction of galaxies in the red distribution is a strong function of local density, increasing from $\approx$10–30% of the population in the lowest density environments, to $\approx$70% at the highest densities. So,they infered that most star-forming galaxies today evolve at a rate which is determined primarily by their intrinsic properties, and independent of their environment, and that the transformation from late to early type must be either sufficiently rapid, or sufficiently rare, to keep the overall color distribution unchanged.

In figure 7-10, we note that the standard deviations of some colors in certain range of local density are abnormally large. We carefully examine colors of each galaxy in the volume-limited sample, and find that above abnormality is due to some galaxy having abnormally large colors

(for example, the galaxy, located at redshift z=0.081, right ascension $RA = 259.2084°$ and declination $DEC = 64.62785°$, has abnormally large g-r color: g-r=7.945).

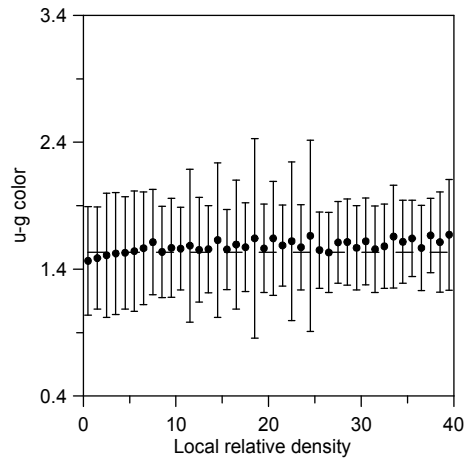

Fig.7　Mean u-g color as a function of the local relative density (within 5th) of galaxies. The dashed line represents the mean u-g color of the volume-limited sample. Error bars are standard deviation in each density bin.

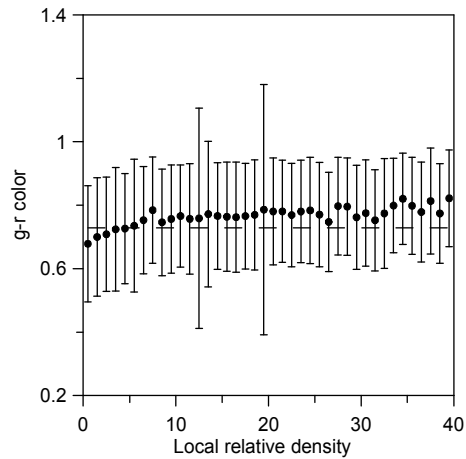

Fig.8　Mean g-r color as a function of the local relative density (within 5th) of galaxies. The dashed line represents the mean g-r color of the volume-limited sample. Error bars are standard deviation in each density bin.

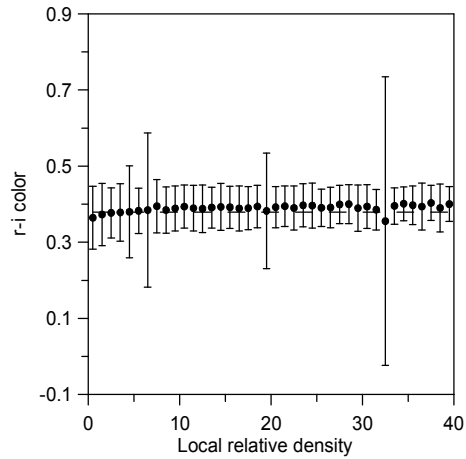

Fig.9  Mean r-i color as a function of the local relative density(within 5th) of galaxies. The dashed line represents the mean r-i color of the volume-limited sample. Error bars are standard deviation in each density bin.

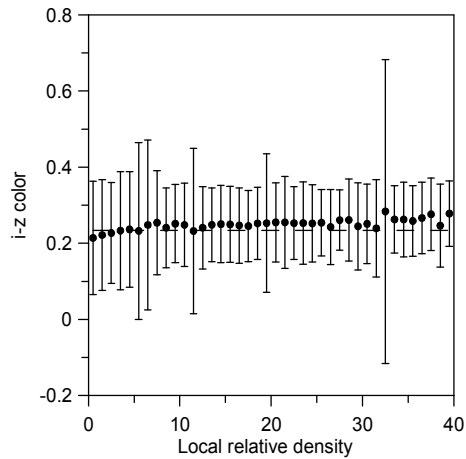

Fig.10  Mean i-z color as a function of the local relative density(within 5th) of galaxies. The dashed line represents the mean i-z color of the volume-limited sample. Error bars are standard deviation in each density bin.

For each galaxy, we also calculate the local three-dimensional galaxy density within the distance to the 10th nearest galaxy. Figure 11 shows the distribution of the local relative density within the distance to the 10th nearest galaxy for all galaxies. Figure 12-18 respectively show the proportion of early type galaxies, the mean luminosity, the mean size, and four optical mean colors as a function of the local relative density of galaxies. As seen from these figures, except that there is a larger scatter in the range of high density, the changes of galaxy properties with the local density within the distance to the 10th nearest galaxy are the same as those with the local density within the distance to the 5th nearest galaxy.

We note that galaxy morphologies strongly depend on local environments: galaxies in dense environments have predominantly early type morphologies, which is confirmed by many other studies. This suggests that in dense environments ther is the existence of the transformation from late to early type. Many physical mechanisms, such as galaxy harassment (Moore et al. 1996), rampressure stripping (Gunn & Gott 1972) and galaxy-galaxy merging (Toomre & Toomre 1972) can explain such process. But we also note that other galaxy properties do not present significant dependence on local environment. Clearly, this puts a important constraint on proposed physical mechanisms

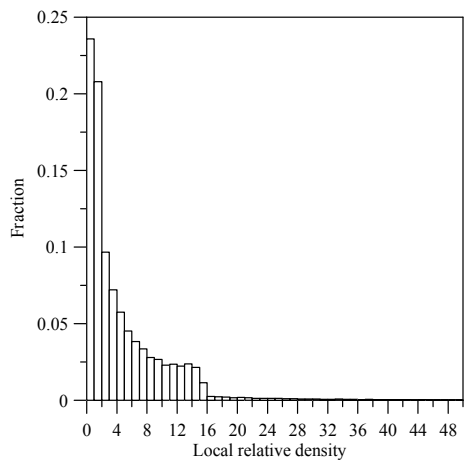

Fig.11   Distribution of the local relative density(within 10 th) for all galaxies.

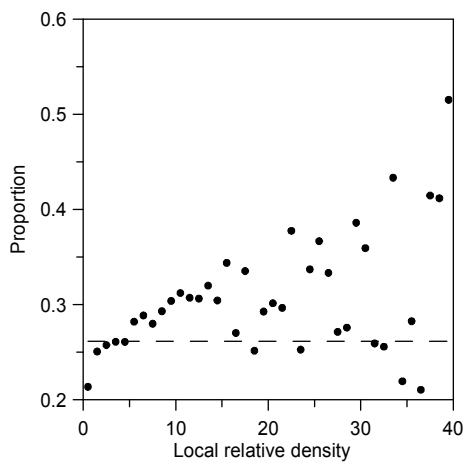

Fig.12 Proportion of early-type galaxies as a function of the local relative density(within 10th) of galaxies. The dashed line represents the proportion of early-type galaxies of the volume-limited sample.

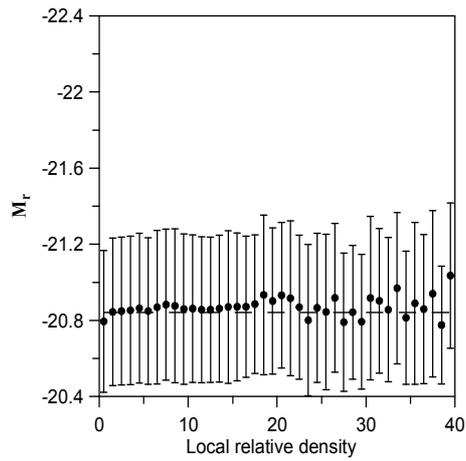

Fig.13 Mean luminosity as a function of the local relative density(within 10th) of galaxies. The dashed line represents the mean luminosity of the volume-limited sample. Error bars are standard deviation in each density bin.

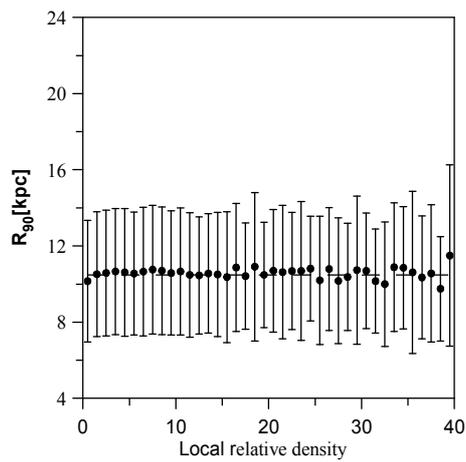

Fig.14 Mean size as a function of the local relative density(within 10th) of galaxies. The dashed line represents the mean size of the volume-limited sample. Error bars are standard deviation in each density bin.

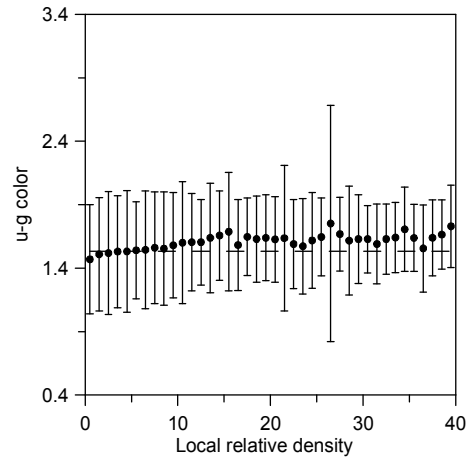

Fig.15  Mean u-g color as a function of the local relative density(within 10th) of galaxies. The dashed line represents the mean u-g color of the volume-limited sample. Error bars are standard deviation in each

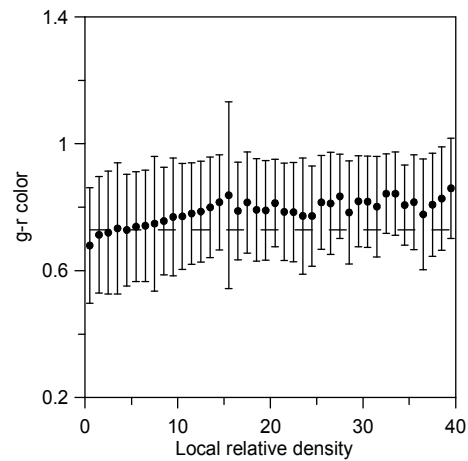

Fig.16  Mean g-r color as a function of the local relative density (within 10th) of galaxies. The dashed line represents the mean g-r color of the volume-limited sample. Error bars are standard deviation in each density bin.

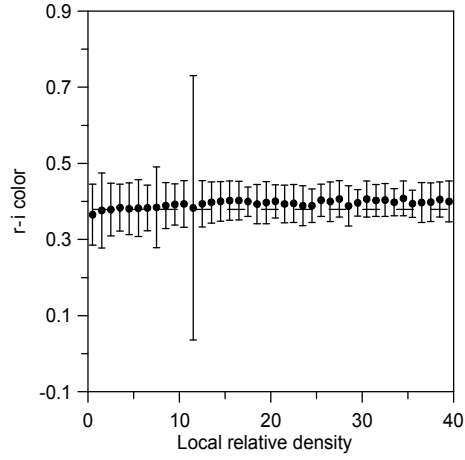

Fig.17 Mean r-i color as a function of the local relative density(within 10th) of galaxies. The dashed line represents the mean r-i color of the volume-limited sample. Error bars are standard deviation in each density bin.

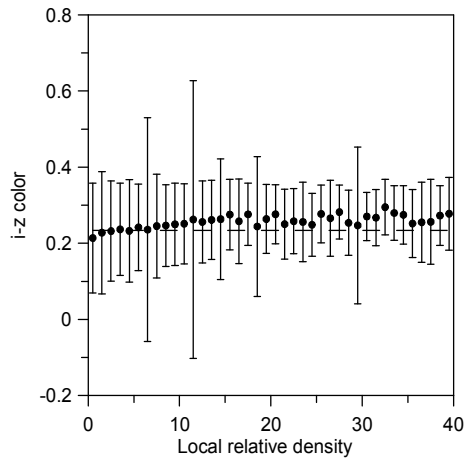

Fig.18 Mean i-z color as a function of the local relative density(within 10th) of galaxies. The dashed line represents the mean i-z color of the volume-limited sample. Error bars are standard deviation in each density bin.

## 4. Summary

In order to investigate the dependence of galaxy properties on local environment, we calculate the local three-dimensional galaxy density within the distance to the 5th nearest galaxy for each galaxy. Because Main galaxy sample is an apparent-magnitude limited sample, we construct a volume-limited sample that extends to $Z_{max} = 0.09$, and and limits the absolute magnitude region: $-22.40 \leq M_r \leq -20.30$, which contains 69381 galaxies, and in which the mean galaxy density is about $2.2386 \times 10^{-3}$ $Mpc^{-3}$. We find that galaxy morphologies strongly depend on local environment: galaxies in dense environments have predominantly early type morphologies, which is expected by many physical mechanisms. But we also note that other galaxy properties do not present significant dependence on local environment. Clearly, this puts a important constraint on proposed physical mechanisms

For each galaxy, we also calculate the local three-dimensional galaxy density within the distance to the 10th nearest galaxy. As seen from figure 12-18, except that there is larger scatter in the range of high density, the changes of galaxy properties with the local density within the distance to the 10th nearest galaxy are the same as those with the local density within the distance to the 5th nearest galaxy.

**Acknowledgements**

Our study is supported by the National Natural Science Foundation of China (10465003).

Funding for the creation and distribution of the SDSS Archive has been provided by the Alfred P. Sloan Foundation, the Participating Institutions, the National Aeronautics and Space Administration, the National Science Foundation, the U.S. Department of Energy, the Japanese Monbukagakusho, and the Max Planck Society. The SDSS Web site is http://www.sdss.org/.

The SDSS is managed by the Astrophysical Research Consortium (ARC) for the Participating Institutions. The Participating Institutions are The University of Chicago, Fermilab, the Institute for Advanced Study, the Japan Participation Group, The Johns Hopkins University, Los Alamos National Laboratory, the Max-Planck-Institute for Astronomy (MPIA), the Max-Planck-Institute for Astrophysics (MPA), New Mexico State University, University of Pittsburgh, Princeton University, the United States Naval Observatory, and the University of Washington.